\documentclass[10pt,twocolumn,showpacs,superscriptaddress,preprintnumbers,floatfix]{revtex4}
\usepackage{latexsym}
\usepackage{amsmath}
\usepackage{amsfonts}
\usepackage{color}
\usepackage{graphicx}

\def\laq{~\raise 0.4ex\hbox{$<$}\kern -0.8em\lower 0.62
ex\hbox{$\sim$}~}
\def\gaq{~\raise 0.4ex\hbox{$>$}\kern -0.7em\lower 0.62
ex\hbox{$\sim$}~}

\def\beq{\begin{equation}}
\def\eeq{\end{equation}}
\def\bea{\begin{eqnarray}}
\def\eea{\end{eqnarray}}
\def\bean{\begin{eqnarray*}}
\def\eean{\end{eqnarray*}}

\def\vp{\varphi}

\def\re {\rangle}
\def\eff{e\!f\!f}

\def \pa {\partial}

\def \b {\beta}
\def \a {\alpha}

\def \ep {\epsilon}

   \def\be{\begin{equation}}
   \def\ee{\end{equation}}
   \def\ba{\begin{eqnarray}}
   \def\ea{\end{eqnarray}}

\begin{document}
\addtolength{\belowdisplayskip}{-0.0cm}
\addtolength{\abovedisplayskip}{-0.0cm}
\title{Tensor Mode Backreaction During Slow-roll Inflation}
\author{G. Marozzi}
\affiliation{Universit\'e de Gen\`eve, D\'epartement de Physique Th\'eorique and CAP,
24 quai Ernest-Ansermet, CH-1211 Gen\`eve 4, Switzerland}
\author{G. P. Vacca}
\affiliation{INFN, sezione di Bologna, via Irnerio 46, I-40126 Bologna, Italy}

\pacs{98.80.Cq, 04.62.+v}
\begin{abstract}
We consider the backreaction of the long wavelength tensor modes produced during a  slow-roll inflationary regime driven by a single scalar field in a spatially flat Friedmann-Lema\^{\i}tre-Robertson-Walker background geometry.
We investigate the effects on non-local observables such as the effective (averaged) expansion rate and equation of state at second order in cosmological perturbation theory.
The coupling between scalar and tensor perturbations induces at second order new tensor backreaction terms beyond the one already present in a de Sitter background. 
We analyze in detail the effects seen by the class of observers comoving with the inflaton field (taken as a clock) and the class of free-falling observers. 
In both cases the quantum backreaction is at least $1/\epsilon$ (with $\epsilon$ the slow-roll parameter) 
larger than the one which can be naively inferred from a de Sitter background.
In particular, we compute the effect for a free massive inflaton model and obtain in both cases a quantum correction
on the background expansion rate of the order of $H^4/(m^2 M_{Pl}^2)$. 
A short discussion on the issue of the breakdown of perturbation theory is given.
 \end{abstract} 
\maketitle

\section{Introduction}
The impact of the quantum backreaction by  tensor and scalar cosmological fluctuations
in an inflationary era has a long history. The effect of two loop infrared gravitons, in a pure 
quantum gravity setting, was conjectured to lead to a secular screening of the effective cosmological
constant \cite{TW}.
A similar conjecture \cite{RHBrev} was then given for the scalar quantum backreaction \cite{MAB1,ABM2}.
On the other hand, this long process has been rather controversial \cite{Uc,All}.
With the aim of settling such controversial issue, 
we have recently performed an series of investigations \cite{FMVV_prl,MV1,MVB,MV_general} based on a new covariant and gauge invariant (GI) approach~\cite{GMV1,GMV2}. In such a context, we have used a GI but observer dependent averaging prescription \cite{GMV1} and a set of covariant and GI effective equations for the averaged geometry \cite{GMV2} (which generalized \cite{Buchert}). 
In particular, we have applied such new GI approach to the study of the scalar quantum backreaction of long wavelength fluctuations considering different classes of observers
\cite{FMVV_prl,MV1,MVB,MV_general}.

Taking advantage of these recent results we evaluate in this paper the tensor mode quantum backreaction during a slow-roll inflationary phase considering two different classes of observers: the one comoving with the inflaton field and the free-falling one.
Such a study is particular appealing for two different reasons. To begin, the recent results  of the BICEP2 collaboration \cite{Ade:2014xna}, 
if confirmed, show an observational imprint of a large production of primordial tensor mode.
Furthermore, tensor backreaction was in the past mainly considered in toy models like a de Sitter background.
On the other hand, as we will show in the following, the use of an ideal de Sitter space-time in the calculation of the tensor mode backreaction is not a good approximation for different reasons. First a de Sitter background does not support scalar perturbations, second it has a series of 
symmetries which are not present in a realistic slow-roll model of inflation.
In particular, the coupling between scalar and tensor perturbations, present in a slow-roll inflationary model at second order in perturbation theory, induces new tensor dependent backreaction terms, beyond the one already present in a de Sitter background, and gives an enhancement of the backreaction effect.

\section{Gauge Invariant Backreaction}
We start by summarizing the approach we are going to follow, which is based on a GI construction of scalar non-local observables.
In general, these observables are constructed performing quantum averages of a scalar field $S(x)$ in a spacetime region which is an hypersurface
$\Sigma_{A_0}=\{x | A(x)=A_0\}$  depending on a second scalar field $A(x)$ with a timelike gradient.
The relative GI averaging prescription $\langle S \rangle_{A_0}$ can be obtained by covariance starting from the (barred) coordinate system 
$\bar{x}^{\mu}=(\bar{t}, \vec{x})$, where the scalar $A$ is homogeneous. Following \cite{GMV1,GMV2} we define
\beq
\langle S \rangle_{A_0}={\langle \sqrt{|\overline{\gamma}(t_0, {\vec{x}})|} 
\,~ \overline{S}(t_0, {\vec{x}}) \re \over  \langle   
\sqrt{|\overline{\gamma}(t_0, {\vec{x}})|} \re}\,,
\label{media}
\eeq
where $\overline{\gamma}(t_0, {\vec{x}})$  is the
determinant of the induced three dimensional metric on $\Sigma_{A_0}$.
Adopting the natural foliation of spacetime associated to the four vector 
\be
n^\mu= -\frac{\partial^\mu A}{(-\partial^\nu A\partial_\nu A)^{1/2}}
\ee
the properties of the class of observers sitting on the hypersurface $A(x)=A_0$ are univocally determined \cite{Mar}.

Let us briefly comment about the non-local nature of the observables considered. 
The expectation values of quantum operators are in general over states which are non local in space, e. g. in a plane wave basis, 
and can be extensively interpreted as a sum over the states (phase space) weighted by the integration volume.
This property is at the base of Eq. (\ref{media}) and leads to the definition of a quantum gauge invariant quantity, which is just dependent on the observer. 
In this sense our observables shares the non-locality of volume averages of late time classical observables.  
Let us also observe that we compute corrections to homogeneous values which are essentially due to long wavelength quantum fluctuations (see below).
Essentially such fluctuations behaves classically, once have crossed the horizon.
Such long wavelength fluctuations have a factorized dependence in space and time so that one can say that the non locality of the observable 
does not affect essentially its time dependence, once its geometrical definition is given.
The definition we employ is the most direct extension of the quantum average over a rigid space translational invariant vacuum,
a case for which volume effect simplifies. Indeed, for the case we consider with a slightly fluctuating volume, we take into account at the perturbative level (second order)  also 
the non trivial fluctuations of the volume encoded in $\sqrt{|\overline{\gamma}(t_0, {\vec{x}})|}$.

On the other hand, it is also possible to consider observables which have a geometrical definition different from the one adopted here. 
For example, one might consider a possibly more physical definition where a class of observers average/measure the contributions of the quantum fluctuations 
on a 2-sphere embedded in their past light-cone (deformed by the fluctuating geometry itself), similarly to the approach developed in \cite{LC-GMNV}.
The sampled quantum fluctuations in the corresponding space-time region are exactly the same in the long wavelength approximation, but this new geometric observable
will have a different dependence in the metric, so that its back-reacted value will present some differences. This interesting analysis will be developed elsewhere.

Going back to the framework associate to Eq.~\eqref{media}, 
the dynamics of a perfect fluid-dominated early Universe can be conveniently described
by an effective scale factor $a_{\eff}=\langle\sqrt{|\bar{\gamma}|}\, \rangle ^{1/3}$.
For the sake of simplicity we choose hereafter $A(x)=t$ at the background level to have standard results neglecting perturbations \cite{Mar}. We can then write a quantum gauge invariant version of the effective
cosmological equation, for the associated expansion rate $H_{eff}= \left(\! \frac{1}{a_{eff}}\frac{\partial \, a_{eff}}{\partial A_0} \!\right)$,  as~\cite{GMV2}

\begin{eqnarray}
\left(\frac{1}{a_{eff}}\frac{\partial \, a_{eff}}{\partial A_0} \right)^2
&=& \frac{8\pi G}{3} \rho_{eff} \label{FirstEffEq} \\
&=&\, \,\frac{1}{9} \left\langle\frac{\Theta}{(-\partial^\mu A\partial_\mu A)^{1/2}} 
\right\rangle_{A_0}^2\,,
\label{genEQ}
\end{eqnarray}
where $\Theta=\nabla_\mu n^\mu$ is the expansion scalar of the timelike congruence $n^\mu$.
In the same way, starting from the following simple relation
\beq
{1\over a_{\eff}}{\pa^2 a_{\eff} \over \pa A_0^2}= 
{\pa \over \pa A_0}\left({1\over a_{\eff}}{\pa a_{\eff} \over \pa A_0}\right)+
\left({1\over a_{\eff}}{\pa a_{\eff} \over \pa A_0}\right)^2 \,,
\label{317}
\eeq
one could define the second effective equation for the averaged geometry as 
 \be
-\frac{1}{a_{\eff}} \frac{\partial^2 \, a_{\eff}}{\partial A_0^2}=   \frac{4\pi G}{3} \left(\rho_{eff}+3 p_{eff} \right)\,.
\label{SecondEffEq}
\ee
Starting from Eqs. (\ref{FirstEffEq}) and (\ref{SecondEffEq}) one can then define an effective equation of state as 
$w_{\eff}=\frac{p_{\eff}}{\rho_{\eff}}$ 
(see, for example, \cite{MVB}).

We shall start from a spatially flat Friedmann-Lema\^{\i}tre-Robertson-Walker (FLRW) background
geometry and consider perturbations up to second order. The metric components of $\{g_{\mu\nu}\}$
are defined as
\bea
g_{00}\!\!&=&\!\! -1\!-\!2 \a\!-\!2 \a^{(2)}\,, \,\,  
g_{i0}=-{a\over2}\!\left(\beta_{,i}\!+\!B_i \right) \!
-\!{a\over2}\!\left(\!\beta^{(2)}_{,i}\!+\!B^{(2)}_i\!\right) 
\nonumber\\
\!\!\!\!g_{ij}\!\! &=&\!\!  a^2 \!\Bigl[ \delta_{ij} \! 
\left( \!1\!-\!2 \psi\!-\!2 \psi^{(2)}\right)
+D_{ij} (E+E^{(2)})
\nonumber\\
& & \!\!\!\!\!\!\!\!\!\!
+{1\over 2} \left(\chi_{i,j}+\chi_{j,i}+h_{ij}\right)
+ {1\over 2} \left(\chi^{(2)}_{i,j}+\chi^{(2)}_{j,i}+h^{(2)}_{ij}\right)\Bigr]
\label{GeneralGauge}
\eea
where $D_{ij}=\partial_i \partial_j- \delta_{ij} (\nabla^2/3)$ and we have removed the upper script for first order quantities. 
We then have that
$\alpha$, $\beta$, $\psi$, $E$ are scalar perturbations, 
$B_i$ and $\chi_i$ are transverse vectors ($\partial^i B_i=0$ and 
$\partial^i \chi_i=0$), and $h_{ij}$ is a traceless and transverse tensor 
($ \partial^i h_{ij}=0=h^{i}_i$). 
 
The Einstein equations connect those fluctuations directly with the matter ones. 
In particular, we consider here a single field inflationary model,
with a self interacting (potential $V$) minimally coupled inflaton scalar field $\Phi$ described by the action
   \be
    S = \int d^4x \sqrt{-g} \left[ 
\frac{R}{16{\pi}G}
    - \frac{1}{2} g^{\mu \nu}
    \partial_{\mu} \Phi \partial_{\nu} \Phi - V(\Phi) \right] \,,
    \label{action}
    \ee
where the inflaton field can be then written to second order as $\Phi(x)=\phi(t)+\varphi(x)+\varphi^{(2)}(x)$.

Clearly the ten degrees of freedom in the metric (\ref{GeneralGauge}) are redundant and a gauge fixing is required, 
which typically removes two scalar and one vector perturbations. 
The cases of interests for us are
the synchronous gauge (SG), defined by $g_{00}=-1$ and $g_{i0}=0$,
the uniform field gauge (UFG), defined by setting $\Phi(x)=\phi(t)$ and 
by another conditions (we consider $g_{i0}=0$), and 
the uniform curvature gauge (UCG), defined by
$g_{ij}=a^2\left[\delta_{ij}+\frac{1}{2} \left(h_{ij}+h^{(2)}_{ij}\right)\right]$.

Let us now give the perturbative expansion (which includes the quantum backreaction) correspondent to Eq.(\ref{genEQ}), namely the expansion rate as seen from an observer sitting on a particular hypersurface.

In the long wavelength (LW) limit we have

\begin{eqnarray} 
& &  \bar{\Theta}=3H-3 H \bar{\alpha}-3\dot{\bar{\psi}}
+\frac{9}{2} H \bar{\alpha}^2+3 \bar{\alpha} \dot{\bar{\psi}}
-6\bar{\psi}\dot{\bar {\psi}}
\nonumber\\
& &\,\,\,\,\,\,\,\,\,\,\,\,\,-3 H \bar{\alpha}^{(2)}-
3 \dot{\bar{\psi}}^{(2)}-\frac{1}{8}h_{ij}\dot{h}^{ij}\,,
\label{Theta}
\end{eqnarray}
\be
\partial_\mu \bar{A} \partial^\mu \bar{A}= 1- 2 \bar{\alpha}+4 
\bar{\alpha}^2- 2 \bar{\alpha}^{(2)}
\label{partA}
\ee
and the measure in the spatial section is given by
\be 
\sqrt{|\bar{\gamma}|}=a^3 \left(1-3 \bar{\psi}+\frac{3}{2}\bar{\psi}^2
-\frac{1}{16} h^{ij}h_{ij}
-3\bar{\psi}^{(2)}\right)\,.
\label{detgamma}
\ee 

Then one has simply to insert Eqs.(\ref{Theta}, \ref{partA}, \ref{detgamma}) in
Eq.(\ref{genEQ})
to obtain 

\be
H_{eff}^2=
\!H^2 \!\left[1\!+\!\frac{2}{H}\langle \bar{\psi}\dot{\bar{\psi}}\rangle \!-\!
\frac{2}{H}\langle \dot{\bar {\psi}}^{(2)}\rangle \!-\! \frac{1}{12H} \langle h_{ij} \dot{h}^{ij}\rangle\right]\,.
\label{EQ1simpl_1}
\ee

Since the quantity above is by definition GI, we can solve the
dynamics of the inflaton and metric fluctuations in any gauge of our choice.
In particular, here we choose to solve the Einstein and matter (inflaton) equations for all the scalar and tensor fluctuations up to second order in the UCG.
This is an extension of the results obtained in \cite{FMVV_II}, where we neglected the tensor fluctuations.
In general, any scalar quantity to second order can be then splitted into two contributions, ${\cal \delta}^{(2)}={\cal \delta}_s^{(2)}+{\cal \delta}_t^{(2)}$, 
one purely scalar ${\cal \delta}_s^{(2)}$ and a second one ${\cal \delta}_t^{(2)}$ induced by the presence of first order tensor modes. 
As it is well known vector modes can be neglected (they die away kinematically) in the inflationary regime.
Therefore we conveniently present the results for just the second order corrections to all the quantities of interest induced by the presence of the tensor fluctuations. The purely scalar induced terms can be found in \cite{FMVV_II}. We find
\begin{equation}
\alpha_t^{(2)} =
\frac{\dot{\phi}}{2H M_{Pl}^2} \varphi_t^{(2)} + s_t \,
\label{alphat}
\end{equation}
\begin{equation}
\frac{H}{a} \nabla^2 \beta_t^{(2)}
= \frac{\dot{\phi}^2}{H M_{Pl}^2} \, \frac{d}{dt} \left
(\frac{H}{\dot{\phi}} \varphi_t^{(2)} \right) - q_t +2 \frac{V}{M_{Pl}^2} s_t.
\label{betat}
\end{equation}

\begin{widetext}

\begin{eqnarray} &&
\ddot{\varphi}_t^{(2)} + 3 H \dot{\varphi}_t^{(2)} - \frac{1}{a^2}\nabla^2 \varphi_t^{(2)} + \left[ V_{\phi \phi} + 2 \frac{d}{dt}\left(3 H +
\frac{\dot H}{H} \right )\right] \varphi_t^{(2)} =
 r_t + \dot \phi \, \dot s_t - 2 V_{\phi} s_t + \frac{\dot \phi}{2 H} \left( q_t - \frac{2 V}{M_{Pl}^2} s_t\right) ,
\label{315}
\end{eqnarray}
where
\begin{eqnarray}
s_t &=&  \frac{1}{8 H}\frac{1}{\nabla^2} \left[\frac{1}{2}\frac{1}{a}
\beta^{, i j} \nabla^2 h_{i j}
+\alpha^{, i j} \dot{h}_{i j}
+\frac{1}{2} h^{k m, i} \dot{h}_{m i, k}- \frac{1}{4} \dot{h}_{i j}
\nabla^2 h^{i j} 
-\frac{3}{4} h_{i j, k}\dot{h}^{i j, k}
-\frac{1}{2} h^{i j}
\nabla^2 \dot{h}_{i j}\right] ,
\label{313}
\end{eqnarray}
\bea &&
q_t = -\frac{H}{2 a} \beta_{,i j}h^{i j}
+\frac{1}{8 a} \beta_{,i j}\dot{h}^{i j}
-\frac{H}{4} h_{i j}\dot{h}^{i j}
-\frac{1}{32} \dot{h}_{i j}\dot{h}^{i j}
+\frac{1}{8 a^2} h^{i j} \nabla^2 h_{i j}
+\frac{3}{32 a^2} h^{i j, k} h_{i j, k}
-\frac{1}{16 a^2} h^{i j, k} h_{i k, j},
\label{314}
\eea
\end{widetext}
\begin{eqnarray}
r_t &=& \frac{\dot{\phi}}{4 a} \beta_{,i j}h^{i j}
-\frac{1}{2 a^2} \varphi_{,i j} h^{i j}
+\frac{\dot{\phi}}{8} h_{i j}\dot{h}^{i j}\,,
\label{317}
\eea
with $M_{Pl}^2=1/(8\pi G)$. Finally, from the divergence-free part of the perturbed $(i,0)$ Einstein
equations, and from the trace-free and divergence-free part of the $(i,j)$
equation, we obtain the evolution equations for the second-order variables
$B_i^{(2)}$ and $h_{ij}^{(2)}$.
However, to second order these variables are decoupled from the scalar
variables $\a^{(2)}$, $\b^{(2)}$, $\vp^{(2)}$, and cannot contribute --
because of their trace-free and transversality properties -- to the
perturbation of the scalar geometric quantities. Hence we shall ignore them.

In the following we want to investigate two classes of observers, the one comoving with the inflaton field and 
the one of the free-falling observers. In these two cases the purely scalar backreaction was found to be zero in the long wavelength
approximation, for any inflaton potential and to all order in the slow-roll approximation~\cite{FMVV_prl, MV_general}. We can then rewrite 
Eq.(\ref{EQ1simpl_1}) as
\be
H_{eff}^2=
\!H^2 \!\left[1\!-\!
\frac{2}{H}\langle \dot{\bar {\psi}}_t^{(2)}\rangle \!-\! \frac{1}{12H} \langle h_{ij} \dot{h}^{ij}\rangle\right]\,,
\label{EQ1simpl_2}
\ee
where non trivial effects must be due to tensor modes.
Therefore we just need to compute the tensor induced  $\bar{\psi}_t^{(2)}$ related to a specific observer.
The coordinate transformation \cite{MetAll}
\beq
x^\mu \rightarrow \bar{x}^\mu= x^\mu + \epsilon^\mu_{(1)} +\frac{1}{2}
\left(\epsilon^{\nu}_{(1)}\pa_\nu \epsilon^{\mu}_{(1)} + \epsilon^{\mu}_{(2)}\right) + \dots
\label{311}
\eeq
induces the following changes which depend on the tensor fluctuations (see, for example, \cite{Mar})
\be
\bar{\varphi}^{(2)}_t=\varphi^{(2)}_t-\frac{\dot{\phi}}{2} \epsilon_{(2)}^0
\ee
\be
\bar{\alpha}_t^{(2)} = \alpha_t^{(2)} - \dot \epsilon^0_{(2)}\quad,\quad
\bar{\beta}_t^{(2)} = \beta_t^{(2)} - \frac{2}{a} \epsilon^0_{(2)} +2 a \dot \epsilon_{(2)},
\label{416}
\ee
\be
\bar{\psi}_t^{(2)}=\psi_t^{(2)}+\frac{H}{2}\epsilon_{(2)}^0+\frac{1}{6}\nabla^2 \epsilon_{(2)}
\,\,,\,\,
\bar{E}_t^{(1)}=E_t^{(2)}-\epsilon_{(2)},
\label{418}
\ee
where we have used the decomposition $\ep_{(2)}^\mu= \left( \ep_{(2)}^0, \pa^i \ep_{(2)}+ \ep_{(2)}^i \right) $. 
For simplicity, in the equations above we have kept only the quantities due to tensor modes and which survive after the quantum averaging procedure. Namely, we have neglected terms where the tensor fluctuations are multiplied by the scalar ones~\cite{Mar}.

\section{Inflaton as a clock}
The observer which sees an homogeneous inflaton field must have $\bar{\varphi}^{(2)}=0$, hence the $\bar{x}^\mu$ reference system will correspond to the  UFG.
Starting from the UCG this condition is obtained by choosing 
\be
\epsilon^0_{(2)}=2 \frac{\varphi_t^{(2)}}{\dot{\phi}}
\quad,\quad
\ep_{(2)}=\int {\rm dt} \left(\frac{1}{a^2} \epsilon^0_{(2)}-\frac{1}{a}\beta_t^{(2)}\right)
\ee

In the  LW limit we then find
\be
H_{eff}^2=
\!H^2 \!\left[1\!-\!
\frac{2}{H}\langle \frac{\rm d}{\rm dt}\left(\frac{H}{\dot{\phi}}\varphi_t^{(2)}\right)\rangle \!-\! \frac{1}{12H} \langle h_{ij} \dot{h}^{ij}\rangle\right]\,.
\label{EQ1UFG}
\ee
In the same limit there is a useful identity which can be derived from Eq.(\ref{betat})
\be
 \frac{\rm d}{\rm dt}\left(\frac{H}{\dot{\phi}}\varphi_t^{(2)}\right)=M_{Pl}^2 \frac{H}{\dot{\phi}^2} \left( q_t-\frac{2 V}{M_{Pl}^2} s_t \right)\,,
 \ee
from which we see that all we need is to compute $\langle q_t \rangle$ and $\langle s_t \rangle$. These quantities receive a non zero contribution just from the quantum averages of forms quadratic in the first order tensor perturbations produced out of the vacuum during the inflationary expansion of the universe.
From Eq.~\eqref{314} one immediately sees, in the LW limit and to the leading order in the slow-roll parameter, that
\be
\langle q_t \rangle=-\frac{H}{4} \langle h_{i j}\dot{h}^{i j}\rangle \,.
\ee

In order to compute $\langle s_t \rangle$, we first note that the first two terms of Eq.~\eqref{313} do not contribute and then
we transform the rest of the expression in a more convenient form by decomposing it in a sum of  symmetrized and antisymmetrized expressions with respect to the permutations of $h_{i j}$ and $\dot{h}_{i j}$.
At the end, we obtain
\be
\langle s_t \rangle=\frac{1}{16H} \langle \frac{ \partial^i\partial_k}{\nabla^2} \left(h^{k m}\dot{h}_{m i}\right)\rangle
-\frac{3}{64H}  \langle h_{i j}\dot{h}^{i j}\rangle\,.
\label{HUFG1}
\ee
The quantum averages in Eq.~\eqref{HUFG1} are computed taking advantage of a standard Fourier decomposition in terms of creation and annihilation operators, amplitudes and polarisation vectors.  The latter, $e_{ij}(\vec{k},s)$, satisfy the normalisation condition 
$\sum_s e^*_{ij}(\vec{k},s) e^{ij}(\vec{k},s)=4$, where the sum in $s$ runs over the two physical polarization state. 
One can show that the average in the first term of Eq.~\eqref{HUFG1} gives $- \langle h_{i j}\dot{h}^{i j}\rangle$ so that we have
\be
\langle s_t \rangle=
-\frac{7}{64H}  \langle h_{i j}\dot{h}^{i j}\rangle\,.
\label{HUFG2}
\ee
Using the results obtained in Eq.~\eqref{EQ1UFG}, and using the relation $\dot{\phi}^2=-2M_{Pl}^2 \dot{H}$ as well as the slow-roll relation $V=3H^2 M_{Pl}^2$,
we find that at leading order in slow-roll approximation
\be
H_{eff}^2=H^2\left[1+\frac{13}{32}\frac{H}{\dot{H}}  \langle h_{i j}\dot{h}^{i j}\rangle\right]\,,
\ee
and the effect of backreaction is enhanced of the order of $1/\epsilon$, with $\epsilon=-\dot{H}/H^2$ the slow-roll parameter, with respect to 
the naive result that one could obtain starting for the de Sitter case \cite{FMVV_GW}.
Let us evaluate it for the free massive inflaton case where $V=\frac{1}{2} m^2 \phi^2$. The graviton dynamics is given by the equation of motion 
\be
\ddot{h}^{i j}+3 H \dot{h}^{i j}-\nabla^2/a^2 h^{i j}=0\,,
\ee
and correspond to the one of a massless test scalar field. 
Then, starting from the results of \cite{Staro-gravitons,starall}, one finds
\be
 \langle h_{i j}\dot{h}^{i j}\rangle=\frac{1}{2}\frac{\rm d}{\rm dt} \langle h_{i j} h^{i j}\rangle=-\frac{3}{\pi^2}\frac{\dot{H} H^3}{m^2 M_{Pl}^2}\,,
\ee
which leads to ($\dot{H}\simeq -m^2/3$)
\be
H_{eff,C}^2\simeq H^2 \left[1-\frac{39}{32\pi^2} \frac{H^4}{m^2 M_{Pl}^2}\right]\,.
\label{BackCom}
\ee
Therefore there is a negative quantum backreaction to the effective expansion rate seen by the observers which use the inflaton as a clock.
The effect of slowing the expansion rate is maximum at the beginning of inflation. For a large initial value of the Hubble factor (which, naively, correspond to a large number of e-folds), the backreaction can not be neglected.

Let us now consider the effective equation of state $w_{eff}=p_{eff}/\rho_{eff}$. In general, 
starting from Eqs. (\ref{FirstEffEq}) to (\ref{SecondEffEq}) 
and assuming the form $H^2_{eff}=H^2 (1+B)$, one has~\cite{MVB}
\be
w_{eff}=-1-\frac{2}{3}\frac{\dot{H}}{H^2}+\left( \frac{\dot{H}}{3H^2} B-\frac{1}{3H} \dot{B}\right)\,.
\ee
Plugging $B=-\frac{39}{32\pi^2} \frac{H^4}{m^2 M_{Pl}^2}$
one finds
\be
w^C_{eff}=-1-\frac{2}{3}\frac{\dot{H}}{H^2}\left(1-\frac{117}{64\pi^2} \frac{H^4}{m^2 M_{Pl}^2}\right)
\ee
which means that for this class of observers the graviton backreaction pushes the inflation in a more de Sitter like phase. 

\section{Geodetic observers}
Let us finally consider a free-falling class of observers, for which the proper time is unperturbed.
The correspondent $\bar{x}^\mu$ reference frame is the synchronous one  ($\bar{\alpha}^{(2)}=0$ together with $\bar{\beta}^{(2)}=0$). 
Similarly to what done in the previous case, these conditions fix the transformation from UCG. One finds $\ep_{(2)}^0= 2 \int {\rm dt} \alpha^{(2)}_t$ and the same as in the previous case for $\ep_{(2)}$.
Then one has
\be
H_{eff}^2\simeq H^2\left[
1\!-\!
2 \langle \alpha^{(2)}_t \rangle -2\frac{\dot{H}}{H}\langle\int \!{\rm dt} \, \alpha^{(2)}_t \rangle  \!-\! \frac{1}{12H} \langle h_{ij} \dot{h}^{ij}\rangle\right].
\label{EQ1GEO}
\ee
From Eq.(\ref{alphat}) we obtain
\be
\langle \alpha^{(2)}_t \rangle=\frac{\dot H}{2 H^2} \int \!{\rm dt} \, \frac{H}{\dot H}\left(\! \langle q_t \rangle-\frac{2V}{M_{Pl}^2}  \langle s_t \rangle \right)+ \langle s_t \rangle
\ee
which in the slow-roll approximation and for a free massive inflaton field reads
$\langle \alpha^{(2)}_t \rangle\simeq-\frac{13}{128\pi^2} \frac{H^4}{m^2M_{Pl}^2}$.

It is then trivial to derive the leading backreaction term in Eq.~\eqref{EQ1GEO}
\be
H_{eff,G}^2 \simeq H^2 \left[1+\frac{39}{160\pi^2}\frac{H^4}{m^2M_{Pl}^2}\right]\,.
\label{EQ1GEO2}
\ee
The "geodesic" class of observers experiences an opposite (positive) backreaction in the effective expansion rate. 

Moving to the effective equation of state for this case, in analogy to the previous case, we derive
\be
w^G_{eff}=-1-\frac{2}{3}\frac{\dot{H}}{H^2}\left(1+\frac{117}{320\pi^2} \frac{H^4}{m^2 M_{Pl}^2}\right)
\ee
which corresponds to a less de Sitter like phase.

The different sign of the graviton backreaction in the two cases above can be traced from the form of the two scalars ($C$ and $G$) which define the two different class of observers (the comoving with the field (UFG) and the "geodesic" ones respectively).
In perturbation theory at second order they were exactly the same when only scalar perturbations were taken into account~\cite{MV_general}.
But after including the tensor perturbations one can write 
$C_t^{(2)}=\varphi^{(2)}_t $ and
\begin{eqnarray}
C_t^{(2)}&=&\frac{1}{\dot{\phi}}\varphi^{(2)}_t  \\
G_t^{(2)}&=&\int {\rm dt} \,\alpha_t^{(2)}\simeq-\frac{1}{\dot{\phi}} \int \frac{{\rm d}H}{H} \varphi_t^{(2)}
\end{eqnarray}
where the last equality is valid in the slow-roll approximation. 
Being the leading dependence of the induced $\varphi_t$ power-like in $H$ with positive exponent, 
it is evident why the corrections to the two scalars are of opposite sign.

\section{Validity of the perturbation theory}
To conclude this analysis let us discuss the regime of validity of the perturbative expansion induced by the tensor fluctuations.
Following \cite{starall}, we evaluate the ratio $\langle \varphi^{(2)}_t \rangle / \phi(t)$ in the UCG, where the fluctuations of the inflaton
field correspond, order by order, to the gauge invariant Mukhanov variable \cite{Mukhanov}.
Considering the previous results one can easily find, at leading order in the slow-roll parameter and in the long wavelength approximation,  that 
\be
\left|\frac{\langle \varphi^{(2)}_t \rangle}{\phi(t)}\right|=\frac{13}{128 \pi^2} \frac{H^4}{m^2 M_{Pl}^2}\,.
\label{BreakTensor}
\ee
Comparing Eq.(\ref{BreakTensor}) with Eqs. (\ref{BackCom}) and (\ref{EQ1GEO2}) one can note that the backreaction effect become important before the ratio in Eq.~\eqref{BreakTensor} becomes of order one.
This breaking, which occurs at the beginning of the inflationary phase, corresponds, for the particular numerical value $M_{Pl}=10^5 m$, to an initial value $H(t_i)~\simeq 10^3 m$.
The regime of validity of the tensor induced fluctuations is different with respect the one of the scalar induced fluctuation studied in \cite{starall}, where
it was shown that the perturbation theory breaks down before the end of inflation, as a consequence of the increase of the ratio 
$\langle \varphi^{(2)}_s \rangle / \phi(t)$, for a value of $H(t_i)$ one order of magnitude lower ($H(t_i)\sim {\cal O}(10^2 m)$).
On the other hand, the ratio $\langle \varphi^{(2)}_s \rangle / \phi(t)$ is zero at the beginning of the inflation and is well below the unit for the major part of the inflationary phase also for value comparable to $H(t_i)\simeq 10^3 m$.
Therefore it is possible to have a non-negligible tensor backreaction effect at the beginning of inflation when both $\langle \varphi^{(2)}_t \rangle / \phi(t)$ and $\langle \varphi^{(2)}_s \rangle / \phi(t)$ are well below the unit and the perturbation theory is under control.

\section{Discussion}
We have shown how spin two quantum fluctuations (tensor modes) affect non-local observables, like the average effective expansion rate and the equation of state, during a cosmological inflationary phase. The computation is based on a covariant approach which leads to GI definitions and depends only on the choice of the observers considered \cite{GMV1,GMV2}. 
The effect seen, which we analyze at second order in perturbation theory, is enhanced of a factor $1/\epsilon$ with respect to the 
one obtained from an adiabatic extension of a de Sitter phase, and is due to the production from the vacuum of linear tensor modes which affect at second order scalar perturbations in the metric and the inflaton field.
Depending on the observer the effect can have a different sign. 
In particular, we have shown that, for a massive free scalar field driven inflation, observers taking the inflaton as a clock see a slowdown in the expansion and a more de Sitter like phase. 
The opposite effect ($5$ times smaller) is experienced by the class of the free-falling observers, which sees the proper time unperturbed.
These effects depend on the inflationary initial condition $H(t_i)$ and are maximum at the beginning of the inflationary phase. The backreaction grows like $H(t_i)^4$ and for the typical value correspondent to a number of e-folds $N$ close to $60$ is negligible (${\cal O}(10^{-7})$). On the other hand, for a large value of $H(t_i)$ this can be non-negligible still when the perturbation theory of the tensor induced and pure scalar perturbations is still reliable.

Let us finally underline that, as mentioned, in the construction of observables there is the possibility of defining new geometrical quantities based on the observation of the fluctuations on null 2-spheres embedded  in the observer past light-cone (see \cite{LC-GMNV}). We shall address this problem in future works.

\vspace{0.2cm}
{\bf Acknowledgements}

We wish to thank Alexei A. Starobinsky for useful discussion.
GM was partially supported by the Marie Curie IEF, Project NeBRiC - "Non-linear effects and backreaction in classical and quantum cosmology".



\begin{thebibliography}{999}
\newcommand{\bb}{\bibitem}


\bibitem{TW}
 N.~C.~Tsamis and R.~P.~Woodard,
  Phys.\ Lett.\  B {\bf 301}, 351 (1993);
  Class.\ Quant.\ Grav.\  {\bf 11}, 2969 (1994);
  Annals Phys.\  {\bf 238}, 1 (1995);
  Annals Phys.\  {\bf 253}, 1 (1997);
  Nucl.\ Phys.\  B {\bf 474}, 235 (1996);
  Phys.\ Rev.\  D {\bf 54}, 2621 (1996);
  Phys.\ Rev.\ D {\bf 80}, 083512 (2009);
  Phys.\ Rev.\ D {\bf 88}, 044040 (2013).

\bibitem{RHBrev}
 R.~H.~Brandenberger,
  hep-th/0210165.

\bibitem{MAB1}
  V.~F.~Mukhanov, L.~R.~W.~Abramo and R.~H.~Brandenberger,
  Phys.\ Rev.\ Lett.\  {\bf 78}, 1624 (1997).
\bibitem{ABM2}
  L.~R.~W.~Abramo, R.~H.~Brandenberger and V.~F.~Mukhanov,
  Phys.\ Rev.\  D {\bf 56} (1997) 3248.

\bibitem{Uc}
  W.~Unruh,
  arXiv:astro-ph/9802323.

\bibitem{All}
L.~R.~W.~Abramo and R.~P.~Woodard,
  Phys.\ Rev.\  D {\bf 60}, 044010 (1999);
  Phys.\ Rev.\  D {\bf 65}, 063515 (2002);
  Phys.\ Rev.\  D {\bf 65}, 043507 (2002).
  G.~Geshnizjani and R.~Brandenberger,
  Phys.\ Rev.\  D {\bf 66}, 123507 (2002);
  JCAP {\bf 04}, 006 (2005);
B.~Losic and W.~G.~Unruh,
  Phys.\ Rev.\  D {\bf 72}, 123510 (2005);
G.~Marozzi,
  Phys.\ Rev.\ D {\bf 76}, 043504 (2007).



\bibitem{FMVV_prl} 
  F.~Finelli, G.~Marozzi, G.~P.~Vacca, G.~Venturi,\\
  Phys.\ Rev.\ Lett.\  {\bf 106}, 121304 (2011)

\bibitem{MV1} 
  G.~Marozzi, G.~P.~Vacca,
  Class.\ Quant.\ Grav.\  {\bf 29}, 115007 (2012).
  
\bibitem{MVB} 
  G.~Marozzi, G.~P.~Vacca, R.~H.~Brandenberger,\\
  JCAP {\bf 1302}, 027 (2013).
  
  
  \bibitem{MV_general} 
  G.~Marozzi and G.~P.~Vacca,
  Phys.\ Rev.\ D {\bf 88}, 027302 (2013).


\bibitem{GMV1}
M. Gasperini, G. Marozzi, and G. Veneziano, JCAP {\bf 03}, 011 (2009).


\bibitem{GMV2}
M. Gasperini, G. Marozzi, and G. Veneziano, JCAP {\bf 02}, 009 (2010).


\bibitem{Buchert} T. Buchert, Gen. Rel. Grav. {\bf 32}, 105 (2000).


\bibitem{Ade:2014xna} 
P.~A.~R.~Ade {\it et al.}  [BICEP2 Collaboration],
  Phys.\ Rev.\ Lett.\  {\bf 112}, 241101 (2014).

\bibitem{Mar}
 G.~Marozzi,
 JCAP {\bf 01}, 012 (2011).

\bibitem{LC-GMNV} 
  M.~Gasperini, G.~Marozzi, F.~Nugier and G.~Veneziano,
  JCAP {\bf 07}, 008 (2011).
  

\bibitem{FMVV_II}
F.~Finelli, G.~Marozzi, G.~P.~Vacca, and G.~Venturi,
Phys.\ Rev.\ D {\bf 69}, 123508 (2004).


\bibitem{MetAll}
  M.~Bruni, S.~Matarrese, S.~Mollerach and S.~Sonego,
  Class.\ Quant.\ Grav.\  {\bf 14}, 2585 (1997).

\bibitem{FMVV_GW}
F.~Finelli, G.~Marozzi, G.~P.~Vacca, and G.~Venturi,
Phys.\ Rev.\ D {\bf 71}, 023522 (2005).


\bibitem{Staro-gravitons} 
A.~A.~Starobinsky,
  JETP Lett.\  {\bf 30}, 682 (1979)
  [Pisma Zh.\ Eksp.\ Teor.\ Fiz.\  {\bf 30}, 719 (1979)].




\bibitem{starall}
F.~Finelli, G.~Marozzi, A. A. Starobinsky, G.~P.~Vacca and G.~Venturi,
Phys. Rev. {\bf D 79}, 044007 (2009);
Phys.\ Rev.\ D {\bf 82}, 064020 (2010).

\bibitem{Mukhanov} 
  V.~F.~Mukhanov,
  JETP Lett.\  {\bf 41}, 493 (1985)
  [Pisma Zh.\ Eksp.\ Teor.\ Fiz.\  {\bf 41}, 402 (1985)].



\end{thebibliography}
\end{document}